\pdfoutput=1
\documentclass[conference,10pt]{IEEEtran}
\IEEEoverridecommandlockouts

\usepackage{cite}
\usepackage{amsmath,amssymb,amsfonts}
\usepackage{graphicx}
\usepackage{textcomp}
\usepackage{xcolor}
\usepackage{booktabs}
\usepackage{multirow}
\usepackage{url}
\usepackage{balance}
\usepackage{enumitem}
\usepackage{subcaption}
\usepackage{eso-pic}

\AddToShipoutPictureBG*{%
  \AtPageLowerLeft{%
    \hspace*{0.75in}%
    \raisebox{0.42in}{%
      \parbox{\dimexpr\paperwidth-1.5in\relax}{\centering\footnotesize
        \copyright~2026 IEEE. Personal use of this material is permitted. Permission from IEEE must be obtained for all other uses, in any current or future media, including reprinting/republishing this material for advertising or promotional purposes, creating new collective works, for resale or redistribution to servers or lists, or reuse of any copyrighted component of this work in other works.
        Accepted for publication in the 44th IEEE International Conference on Computer Design (ICCD), 2026.}}}}

\begin{document}

\title{\LARGE OpenULF: A Modular Toolkit Enables Data-Driven Ultra-Low-Field MRI with both Classical and Quantum Sensors}

\author{%
\IEEEauthorblockN{Hanhan Wu, Yuchen Yuan, Weiwen Jiang}
\IEEEauthorblockA{\textit{George Mason University}, Fairfax, VA, USA\\
hwu28@gmu.edu, yyuan21@gmu.edu, wjiang8@gmu.edu}
}

\maketitle

\begin{abstract}
Ultra-low-field (ULF) Magnetic Resonance Imaging (MRI) offers a portable, low-cost alternative to conventional scanners by operating at field strengths below 10\,mT. At these fields, two sensor technologies compete: quantum SQUID (Superconducting Quantum Interference Device) sensors, which provide high sensitivity but require cryogenic cooling, and classical induction coils, which are inexpensive and room-temperature but produce weaker signals. 
With conventional physics-based imaging, both suffer from low signal-to-noise ratios (SNR). Data-driven reconstruction has overcome similar limits in related inverse problems, but its use for ULF-MRI has been hindered by a lack of open datasets, benchmarks, and flexible simulators.
To close this gap, we present OpenULF, a modular, open-source simulation toolkit that generates paired training data and supports end-to-end machine-learning (ML) reconstruction. OpenULF integrates a Bloch-equation-validated physics simulator, calibrated SQUID and coil receiver models, reconstruction ML networks, and an automated design-space exploration engine. Architecturally, a receiver abstract base class, a typed HDF5 contract that decouples physics from learning, and a resumable sweep engine make sensors, models, and datasets independently substitutable. Using OpenULF, we train 774 models across four anatomical datasets and conduct three design-space explorations.
The effectiveness of ML reconstruction depends strongly on the dataset and the sensor. We identify a SQUID-to-coil crossover near 450\,$\mu$T and demonstrate that sensor hardware and ML model capacity can be traded off during system design. Sensor and reconstruction choices must therefore be co-designed: the best sensor is set by the dataset and field strength, not by SNR.

\end{abstract}


\section{Introduction}
\label{sec:intro}
Ultra-low-field (ULF) MRI operates at field strengths below 10\,mT and is a portable, low-cost alternative to conventional 1.5--3\,T scanners~\cite{vesanen2013, zotev2008, hyperfine2020}. ULF systems can be built from lightweight permanent magnets rather than multi-ton superconducting coils, enabling deployment where conventional MRI is impractical~\cite{hyperfine2020, sheth2021}. At these field strengths, two fundamentally different sensor technologies are available.
Quantum sensors based on Superconducting Quantum Interference Devices (SQUIDs) provide high sensitivity independent of frequency but require cryogenic cooling with liquid helium.
Classical induction coils are inexpensive and operate at room temperature, but their signal scales with the Larmor frequency $\omega_0 = \gamma B_0$ and is therefore very weak at ULF. With conventional physics-based reconstruction, both sensing modalities suffer from low signal-to-noise ratios (SNR), resulting in limited image quality. In our simulation, even the SQUID achieves only about 11\,dB in Peak Signal-to-Noise Ratio (PSNR) and 0.16 in Structural Similarity Index (SSIM) on brain data, well below diagnostic quality.

Recent progress in data-driven reconstruction for inverse problems, including ultrasound computed tomography (USCT)~\cite{usct_dl2020, yuan2026muxusctnoiserobustneuralnetwork} and full-waveform inversion (FWI)~\cite{fwi_dl2023, yuan2026twoteachers}, suggests that machine learning (ML) can overcome the severe SNR limitations of ULF-MRI by learning to recover anatomical structure from noisy measurements. However, systematic exploration of data-driven ULF-MRI has been hindered by three gaps: the lack of open datasets, the absence of standardized benchmarks, and the shortage of flexible simulation frameworks. Existing MRI simulators such as MRiLab~\cite{mrilab} and JEMRIS~\cite{jemris} target clinical field strengths. Their receiver models do not support SQUID sensors or prepolarization, and their noise models assume sample-noise-dominated regimes that do not apply at ULF. ML reconstruction frameworks such as fastMRI~\cite{fastmri} assume fixed, pre-acquired clinical-quality data. Table~\ref{tab:tools} summarizes the feature coverage. No existing tool integrates ULF-specific sensor physics, ML reconstruction, and automated parameter sweeps in a single pipeline.

\begin{table}[t]
\centering
\caption{Feature comparison of MRI simulation tools
  (\checkmark\,=\,supported, \textsf{--}\,=\,not). ``Automated param.\
  sweep'' means a built-in resumable, deduplicated sweep engine, not the
  general scripting that MRiLab and JEMRIS also provide; ``Multi-model
  library'' means bundled reconstruction models.}
\label{tab:tools}
\small
\setlength{\tabcolsep}{2.5pt}
\begin{tabular}{l|cccc}
\toprule
Feature & MRiLab & JEMRIS & fastMRI & OpenULF \\
\midrule
Bloch spin dynamics   & \checkmark & \checkmark & \textsf{--} & \checkmark \\
Pulse sequence sim.   & \checkmark & \checkmark & \textsf{--} & \checkmark \\
ULF sensor models     & \textsf{--} & \textsf{--} & \textsf{--} & \checkmark \\
SQUID + Coil in one   & \textsf{--} & \textsf{--} & \textsf{--} & \checkmark \\
ML reconstruction     & \textsf{--} & \textsf{--} & \checkmark & \checkmark \\
Multi-model library   & \textsf{--} & \textsf{--} & \checkmark & \checkmark \\
Automated param.\ sweep & \textsf{--} & \textsf{--} & \textsf{--} & \checkmark \\
Modular composition   & \textsf{--} & \textsf{--} & \textsf{--} & \checkmark \\
\bottomrule
\end{tabular}
\end{table}

\begin{figure*}[t]
  \centering
  \includegraphics[width=0.97\textwidth]{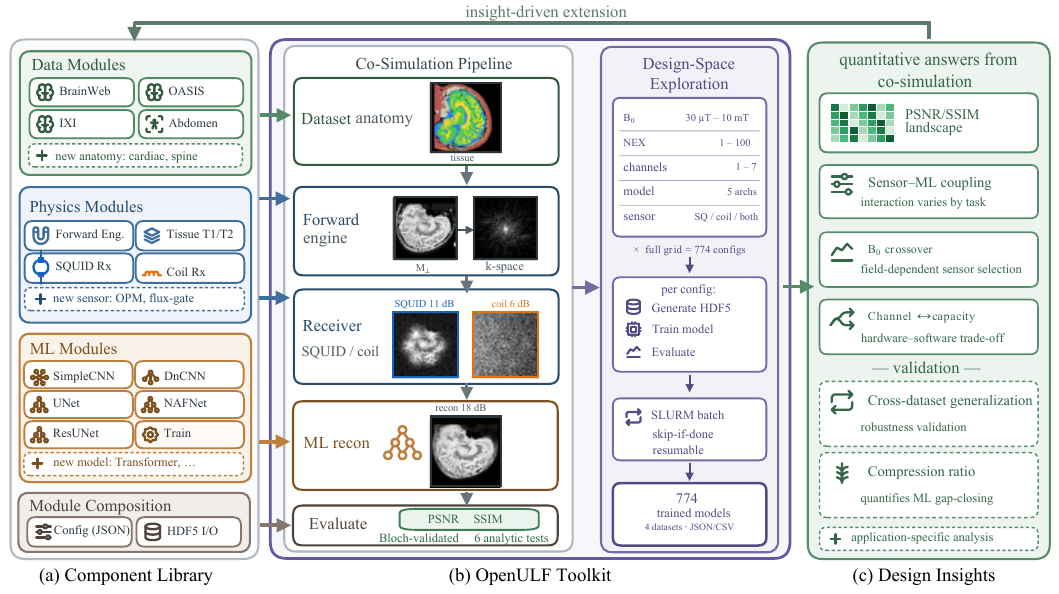}
  \caption{Overview of OpenULF. (a)~Component library with extensible physics, ML, data, and composition modules. (b)~OpenULF toolkit: the co-simulation pipeline (left sub-box) transforms tissue maps through sensor-specific receivers and ML reconstruction; the design-space exploration engine (right sub-box) sweeps $B_0 \times \text{NEX} \times \text{channels} \times \text{model} \times \text{sensor}$, training 774 models across four datasets. (c)~Design insights: three quantitative design rules plus cross-dataset and compression-ratio validation. The top-level feedback arrow indicates that insights drive iterative extension of the component library.}
  \label{fig:platform}
\end{figure*}

To address these gaps, we present OpenULF,\footnote{Code available at \url{https://github.com/Wu-Hanhan/OpenULF}.} a modular, open-source simulation toolkit for data-driven ULF-MRI with both classical and quantum sensors (Fig.~\ref{fig:platform}). OpenULF integrates a Bloch-equation-validated physics simulator ($\le$0.01\% error across six analytical tests), calibrated SQUID and coil receiver models with ULF-specific noise, five reconstruction networks spanning 0.2\,M to 10\,M parameters, and an automated design-space exploration engine. 
This work makes three contributions:
\begin{itemize}[leftmargin=*,nosep]
\item A modular co-simulation architecture in which a receiver abstraction and a typed on-disk contract decouple calibrated ULF sensor physics (both SQUID and coil receivers, prepolarization, and sensor-noise-dominated regimes) from end-to-end ML reconstruction, so researchers can substitute their own architectures or anatomical datasets without modifying the physics simulator. To the best of our knowledge, no existing tool provides this combination.
\item An automated sweep engine that explores the $B_0 \times \text{NEX} \times \text{channels} \times \text{model} \times \text{sensor}$ design space at scale, training 774~models across four anatomical datasets (BrainWeb, OASIS, IXI, and a synthetic abdomen set).
\item Three quantitative design rules, derived from systematic experiments, that reveal interactions between sensor configuration and ML reconstruction not available from component-level metrics such as SNR.
\end{itemize}

Our experiments produce three main observations. 
First, the effect of ML reconstruction on the SQUID-coil quality gap depends on both the dataset and the sensor: ML compresses the gap on some anatomies and amplifies it on others. Second, a SQUID-to-coil crossover near 450\,$\mu$T separates a SQUID-favored low-field regime from a coil-favored high-field one. Third, sensor channel count and ML model capacity are interchangeable design variables. None of these follow from isolated SNR analysis; they require running sensor, acquisition, and ML choices together end-to-end.

The remainder of this paper provides background and motivation (Sections~\ref{sec:background} and~\ref{sec:motivation}), the OpenULF platform and its validation (Section~\ref{sec:platform}), and the three design-space explorations with their resulting design rules (Section~\ref{sec:experiments}).
\section{Background and Related Work}
\label{sec:background}

\subsection{Ultra-Low-Field MRI}

Clinical MRI operates at 1.5--3\,T using superconducting magnets that
weigh several tons and cost millions of dollars. Ultra-low-field MRI
reduces $B_0$ to below 10\,mT, typically 10--200\,$\mu$T, trading
signal strength for portability and cost~\cite{vesanen2013,
zotev2008}. ULF-MRI brain imaging has been demonstrated at national labs and in
commercial scanners such as Hyperfine's Swoop~\cite{zotev2009,
vesanen2013, hyperfine2020}. Because the equilibrium
magnetization is proportional to $B_0$, a 50\,$\mu$T scanner produces
a signal roughly 60{,}000 times weaker than a 3\,T system.
Prepolarization partially compensates for this deficit: the sample is
magnetized at a stronger field $B_\text{prepol} \gg B_0$ before
readout at $B_0$~\cite{vesanen2013}.

Two fundamentally different sensor technologies compete at these field
strengths. Both detect the precessing transverse magnetization
$M_\perp$ through the reciprocity flux $\Psi$
(Eq.~\ref{eq:reciprocity}), but they convert that flux into a
measurable voltage differently.

SQUID sensors (Superconducting Quantum Interference Devices) sense
magnetic flux via the Josephson effect. When operated in a flux-locked
loop (FLL), the output voltage is
\begin{equation}
  V_\text{SQUID} = V_\Phi \cdot \Psi
  \label{eq:squid}
\end{equation}
where $V_\Phi$\,[V/Wb] is the system flux-to-voltage coefficient. The
key property is that $V_\Phi$ does not depend on frequency: a SQUID
responds equally well at 2\,kHz ($B_0 = 50\,\mu$T) and at 128\,MHz
($B_0 = 3$\,T). This makes SQUIDs well suited to the very low Larmor
frequencies of ULF-MRI, but the Josephson junction must be kept below
its critical temperature, which requires a liquid-helium dewar.

Induction coils detect the time derivative of flux via Faraday's law:
\begin{equation}
  \mathcal{E}_\text{coil} = -N\,\frac{d\Psi}{dt}
  \approx -N\,\omega_0\,\Psi
  \label{eq:coil}
\end{equation}
where $N$ is the number of turns and $\omega_0 = \gamma B_0$ is the
Larmor frequency. The induced EMF grows linearly with $B_0$, so at
clinical fields coils are the standard sensor; but at $B_0 =
50\,\mu$T the EMF falls by the same factor, placing the coil at a
severe disadvantage. The benefit is that coils operate at room
temperature and require no cryogenic infrastructure.

This split in the physics, frequency-independent gain for the SQUID
versus $\omega_0$-proportional gain for the coil, is the root of the
sensor design problem that OpenULF addresses.

\subsection{ML for MRI Reconstruction}

Deep learning has become widely used in MRI image reconstruction, especially at low SNR where classical methods such as Wiener filtering, BM3D, and total variation underperform~\cite{lundervold2019}. UNet~\cite{unet2015} and DnCNN~\cite{dncnn2017} are widely used; more recent architectures include NAFNet~\cite{nafnet2022} and diffusion-based reconstructors~\cite{song2022}. Compressed sensing approaches~\cite{lustig2007} reduce acquisition time but still require adequate input SNR. At the noise levels typical of ULF-MRI (input SNR roughly 5--11\,dB), ML reconstruction can recover several dB of PSNR depending on model capacity, making it increasingly important for ULF-MRI.

\subsection{Related Simulation Tools}

Several MRI simulation frameworks exist. MRiLab~\cite{mrilab} and JEMRIS~\cite{jemris} simulate spin dynamics and k-space acquisition with configurable pulse sequences, but they do not include an ML reconstruction stage or sensor noise models calibrated for ULF-MRI. Their receiver models target clinical RF frequencies in sample-noise-dominated regimes and do not support SQUID sensors or prepolarization, both of which are specific to ULF-MRI. Virtual Scanner~\cite{virtualscanner2019} is a browser-based educational simulator; BART~\cite{bart2016} and SigPy~\cite{sigpy2020} are reconstruction libraries that operate on pre-acquired k-space; and fastMRI~\cite{fastmri} provides benchmark data and ML baselines but assumes fixed, clinical-quality input. As summarized in Table~\ref{tab:tools}, none integrates sensor-specific physics, ML reconstruction, and automated parameter sweeps in a single pipeline.
\section{Motivation}
\label{sec:motivation}

Designing a portable ULF-MRI system requires joint decisions across sensor hardware, acquisition protocol, and reconstruction software. Three observations motivate the need for an integrated simulation toolkit.

Physics simulators and ML frameworks are developed in isolation: physics tools such as MRiLab and JEMRIS stop at signal generation, while ML frameworks such as fastMRI assume pre-acquired data and never model the sensor physics that produced it. Combining them is not a one-time cost but a recurring one: every new sensor-or-model pairing means re-bridging data formats, coordinate systems, and noise models by hand, with glue code rewritten for every study and never reused. OpenULF instead fixes that boundary once, as a typed on-disk contract, so any sensor composes with any model.

Sensor SNR does not reliably predict post-ML reconstruction quality.
The higher-SNR sensor is conventionally assumed to give the better
image, but ML is a non-linear transform whose outcome depends on model
capacity: as Section~\ref{sec:experiments} shows, it can shrink or widen
the sensor gap depending on the dataset, so SNR-based selection is
unreliable once ML is in the loop.

The design space is also combinatorially large. A ULF-MRI system is parameterized by $B_0$, NEX (number of signal averages), SQUID array channel count, and ML model
choice. A modest grid of 13~$B_0$ values, 8~NEX settings, 7~channel
counts, 5~models, and 3~sensor configurations already exceeds
10{,}000 design points. Automated exploration is needed to extract design rules from such a space.
\section{OpenULF Platform}
\label{sec:platform}

OpenULF is organized into three parts. The component library
(Section~\ref{ssec:library}) provides reusable physics, ML, and
data-generation components that can be combined without writing custom
integration code. The end-to-end co-simulation pipeline
(Section~\ref{ssec:pipeline}) runs these modules in sequence for any
given sensor, model, and dataset configuration, producing
reconstruction quality metrics rather than intermediate SNR values.
The automated sweep engine (Section~\ref{ssec:sweep}) runs the
pipeline across large parameter grids, making it practical to explore
thousands of design points. Architecturally, OpenULF rests on three
reusable mechanisms, none of them MRI-specific: (i)~a receiver abstract
base class that reduces a new sensor to five enforced extension points
over a shared acquisition core; (ii)~a typed HDF5 contract that acts as
a compiler-style intermediate representation between a physics
front-end with no ML dependencies and a learned back-end, so the two
halves are independently substitutable; and (iii)~a make-like,
idempotent sweep engine that turns design-space exploration into a
resumable, deduplicated build. This work's two receivers, five
reconstruction networks, and four dataset generators were each added
through these interfaces alone, without modifying the shared acquisition
core, the HDF5 contract, or the sweep engine.

\subsection{Component Library}
\label{ssec:library}

The component library provides physics, ML, and data modules, wired
together by a thin composition layer; all four are shown on the left of
Fig.~\ref{fig:platform}.

\subsubsection{Physics Modules}

The forward engine computes transverse magnetization $M_\perp$\,[A/m]
for each tissue class using a three-phase prepolarization
model~\cite{vesanen2013}. The sample is first polarized at
$B_\text{prepol}$ for duration $t_\text{pol}$, then transferred to
$B_0$ with a delay $t_\text{xfer}$, and finally excited by a
90$^\circ$ pulse for spin-echo readout. The magnetization available
before excitation is
\begin{equation}
  M_\text{avail} = M_\text{eq}(B_0) +
    \bigl(M_\text{pol} - M_\text{eq}(B_0)\bigr)\,
    e^{-t_\text{xfer}/T_1(B_0)}
  \label{eq:prepol}
\end{equation}
where $M_\text{pol} = M_\text{eq}(B_\text{prepol})\,(1 -
e^{-t_\text{pol}/T_1(B_\text{prepol})})$. Prepolarization is essential
for ULF-MRI: without it, the equilibrium magnetization at
$B_0 = 50\,\mu$T is too weak for imaging. With
$B_\text{prepol} = 0.1$\,T, the available magnetization increases by
roughly three orders of magnitude, bringing the signal into the
detectable range for both sensor types. Tissue-specific $T_1$, $T_2$,
and proton density values are drawn from the BrainWeb
atlas~\cite{brainweb1998} with field-dependent relaxation following the
Bottomley model~\cite{Bottomley1984}. The engine converts the $M_\perp$ image into k-space
via 2-D FFT.

The receiver layer is the architectural core of OpenULF. Both sensors
are concrete subclasses of a single abstract base class,
\texttt{ReceiverBase}, which provides the shared acquisition path
(reciprocity-based flux integration~\cite{hoult1976}, NEX signal
averaging, ADC quantization, GPU/CPU-portable colored-noise synthesis,
and root-sum-of-squares multi-coil combination) in one core:
\begin{equation}
  \Psi = \int \frac{\mathbf{B}_1(\mathbf{r})}{I_0}
         \cdot M_\perp(\mathbf{r})\,d^3r
  \quad [\text{Wb}]
  \label{eq:reciprocity}
\end{equation}
A concrete sensor supplies only five hooks: the per-unit-current
sensitivity map $\mathbf{B}_1/I_0$ and a per-readout voltage transform,
plus three frequency responses (flux-to-voltage gain, transfer function
$H(f)$, and noise power spectral density). The two shipped sensors
differ \emph{only} in these hooks: the room-temperature induction coil
(Faraday EMF, Johnson--Nyquist noise) and the cryogenic SQUID
gradiometer (frequency-independent gain, fT-scale noise
floor~\cite{clarke2004, zotev2008}). Both reuse the same acquisition
base, so a new sensor touches only the five hooks, never the
acquisition path.
Adding a heterogeneous sensor thus means implementing one class
against this interface,
leaving the rest of the pipeline untouched; multi-channel arrays are
produced by a factory that arranges $N$ hexagonal elements with
decorrelated noise. Per-sensor geometry (the SQUID in a dewar, the coil on the
scalp) is configuration, not a
hard-wired assumption~\cite{zotev2008}.

\subsubsection{ML Modules}

OpenULF includes five reconstruction architectures that span nearly two
orders of magnitude in parameter count: SimpleCNN (0.22\,M, 8-layer
plain CNN), DnCNN (0.56\,M, 17-layer denoiser~\cite{dncnn2017}), UNet
(7.76\,M, encoder-decoder with skip
connections~\cite{unet2015}), ResUNet (10.3\,M, UNet with residual
blocks), and NAFNet (7.4\,M, non-linear activation free
network~\cite{nafnet2022}). All expose the same single- or dual-channel
input (SQUID, coil, or both) to single-channel magnitude output
interface and share a unified training loop; a new architecture is
added by dropping in
a module that honors this interface, with no change to the data or
physics layers. Spanning capacity this widely makes the model a tunable
design resource in its own right, echoing model-compression work that
trades network size against quantum and embedded hardware
budgets~\cite{hu2022qnncompression}.

\subsubsection{Data Modules}

Four dataset generators produce paired (noisy input, ideal target)
HDF5 files from different anatomical sources: BrainWeb (20 fuzzy
phantoms, 1{,}200 training slices), OASIS (414 subjects,
11{,}838 slices), IXI (583 subjects, 12{,}835 slices), and a synthetic abdomen
set (500 phantoms). A new anatomy is added by writing a generator that
emits the same (\texttt{ideal}, \texttt{squid}, \texttt{coil}) HDF5
contract; every generator is parameterized by $B_0$, NEX,
$B_\text{prepol}$, and array channel count (1--7 hexagonal SQUID
channels~\cite{zotev2007}), which enables the design-space sweeps in
Section~\ref{sec:experiments}.

\subsubsection{Composition}

A thin composition layer wires these modules into runs without custom
integration code: a configuration interface (command-line parameters,
captured per run as JSON manifests) and a typed HDF5 I/O contract: the
(\texttt{ideal}, \texttt{squid}, \texttt{coil}) arrays plus the physics
attributes that produced them. This layer is what makes the physics and
learning halves composable and independently substitutable, as detailed
next.

\subsection{End-to-End Co-Simulation Pipeline}
\label{ssec:pipeline}

The co-simulation pipeline chains the full signal path (tissue atlas,
forward engine, receiver, ML reconstruction, and evaluation via
PSNR/SSIM) for any $(\text{sensor}, \text{model}, \text{dataset})$
configuration. The HDF5 dataset is the contract between the two halves
of the toolkit: physics modules emit the arrays (\texttt{ideal},
\texttt{squid}, \texttt{coil}) with per-image normalization factors,
and the ML modules import nothing from the physics layer, consuming
only this file. Because the boundary is a stable on-disk schema rather
than a shared in-memory API, the physics and learning stages evolve
independently: a new architecture never touches the simulator, and a
trained checkpoint can be re-evaluated against a regenerated dataset.
A run is composed from a small set of command-line parameters (sensor,
model, dataset, and physics settings); each dataset records its full
physics configuration as HDF5 attributes and each training run emits a
JSON manifest of hyperparameters, parameter count, and hardware,
giving end-to-end provenance from physics parameter to trained
checkpoint.

As its end-to-end metric the pipeline reports the compression
ratio, which measures how much of the physical quality difference
between sensors survives ML reconstruction. This metric cannot be
recovered from SNR analysis and is defined in full in
Section~\ref{sec:experiments}.

\subsection{Automated Parameter-Sweep Engine}
\label{ssec:sweep}

The sweep engine turns the co-simulation pipeline into the inner loop
of an automated design-space exploration over $B_0 \times \text{NEX}
\times \text{channels} \times \text{model} \times \text{sensor}$. It
generates a parameterized HDF5 dataset per physics point and submits
one cluster job per $(\text{model}, \text{sensor})$ training run. The
sweep behaves like a build system: datasets and checkpoints are targets
guarded by existence checks, so a preempted cluster job requeues and
resumes where it stopped, and design points shared across studies are
deduplicated rather than retrained. Per-run logs are then collected into a
single index keyed by all sweep dimensions. Rather than enumerate the
full Cartesian product (over 10{,}000 points), each of the three studies
fixes all axes but one and sweeps that axis at full resolution, so the
774 trained models are targeted, representative slices rather than
exhaustive coverage: Study~1 varies sensor, model, and dataset at the
default field; Study~2 sweeps $B_0$ (13 values, 30--10{,}000\,$\mu$T) and
NEX (8 settings, 1--100); and Study~3 sweeps channel count (1--7). Design
points shared across studies are trained once and reused.

\subsection{Validation}

We check the physics simulator in two complementary ways: numerical
correctness and physical grounding. For correctness,
Table~\ref{tab:validation} compares the forward engine against
closed-form Bloch-equation solutions over six tests, all at or below
0.01\% relative error, and an independent cross-check against the EPG simulator
\texttt{mri-sim-py}, a different
simulation method, gives consistent results. Near-zero error here
is exactly what a correct implementation should produce: it confirms
the equations are solved correctly. Physical realism is a separate
question, answered by calibration that grounds every physical
parameter ($B_0$, receiver geometry, noise floors, standoff distances,
and $T_1/T_2$ field dependence) in published
measurements~\cite{vesanen2013, zotev2008, clarke2004, Bottomley1984}.

\begin{table}[t]
\centering
\caption{Numerical-correctness check of the forward engine against
  closed-form Bloch solutions. Near-zero error is expected of a correct
  implementation; physical realism comes from measured parameters
  (Section~\ref{ssec:library}). Here 0.00\% denotes error below the
  two-decimal display threshold.}
\label{tab:validation}
\small
\setlength{\tabcolsep}{3pt}
\begin{tabular}{clc}
\toprule
\# & Test description & Error \\
\midrule
1 & Single-voxel $M_\perp$ vs.\ spin-echo (6 tissues) & 0.00\% \\
2 & $T_1$ contrast: GM/WM ratio vs.\ TR (5 pts) & 0.00\% \\
3 & $T_2$ decay: signal vs.\ TE (6 points) & 0.00\% \\
4 & Prepolarization (12 config combinations) & $<$0.01\% \\
5 & Image: Parseval, k-space DC, tissue order & 0.00\% \\
6 & $T_1$ field dependence~\cite{Bottomley1984} & 0.00\% \\
\bottomrule
\end{tabular}
\end{table}

\section{Experiments}
\label{sec:experiments}

\subsection{Experimental Setup}

We use four anatomical datasets generated by the sweep engine:
BrainWeb (20 fuzzy phantoms, 1{,}200/171/336 train/val/test slices),
OASIS\footnote{Data were provided in part by OASIS-1 (Cross-Sectional;
Principal Investigators D.~Marcus, R.~Buckner, J.~Csernansky, J.~Morris;
grants P50~AG05681, P01~AG03991, P01~AG026276, R01~AG021910, P20~MH071616,
U24~RR021382)~\cite{oasis2007}.} (414 subjects, 11{,}838/1{,}684/3{,}458),
IXI\footnote{The IXI data were provided by the Biomedical Image Analysis
Group, Imperial College London, under CC~BY-SA~3.0~\cite{ixi}.} (583
subjects, 12{,}835/1{,}831/3{,}728), and a synthetic abdomen set (500 phantoms,
350/50/100). For OASIS and IXI the splits are subject-level; for BrainWeb and the
synthetic abdomen set they are phantom-level. No subject or phantom
contributes slices to both training and test, ruling out same-source
leakage. All
images are $128 \times 128$ and normalized per-image to $[0,1]$ using
the 99.5th percentile. The default physics
configuration follows Vesanen et al.~\cite{vesanen2013}: $B_0 =
50\,\mu$T, $B_\text{prepol} = 0.1$\,T, TE\,=\,50\,ms, NEX\,=\,10,
7-channel hexagonal SQUID array whose elements are root-sum-of-squares
combined into one input image. Five architectures are trained with
MSE loss, Adam ($\text{lr} = 10^{-3}$, weight decay 0), batch size 16,
early stopping (patience 20), and standard augmentation. Each model is trained separately for
SQUID-only (1-ch), coil-only (1-ch), and both (2-ch) inputs. All runs
use A100 GPUs. The sweep engine trained 774 models in
total.

\subsection{Sensor-ML Coupling}
\label{ssec:study1}

Can ML reconstruction close the quality gap between the expensive
SQUID and the cheap coil, and is its effect on that gap consistent
across anatomies? We answer this by running all five models on both
sensors across all four datasets and reading reconstruction quality
end-to-end rather than from input SNR.

\begin{table}[t]
\centering
\caption{Test-set PSNR (dB) and SSIM across 4 datasets, 5 models, and
  3 sensor configurations. The Input row reports the raw sensor
  output (before ML) against the ideal image. Gap = SQUID $-$ Coil in PSNR.}
\label{tab:main}
\small
\setlength{\tabcolsep}{2.2pt}
\begin{tabular}{ll|cc|cc|cc|r}
\toprule
 & & \multicolumn{2}{c|}{SQUID} & \multicolumn{2}{c|}{Coil} & \multicolumn{2}{c|}{Both} & \\
Dataset & Model & dB & SSIM & dB & SSIM & dB & SSIM & Gap \\
\midrule
\multirow{6}{*}{\rotatebox{90}{\scriptsize BrainWeb}}
 & Input     & 11.0 & .16 &  5.3 & .02 & -- & -- & +5.8 \\
 & SimpleCNN & 18.0 & .41 & 12.2 & .16 & 18.1 & .49 & +5.9 \\
 & DnCNN     & 18.9 & .66 & 13.8 & .37 & 19.0 & .68 & +5.2 \\
 & UNet      & 19.8 & .80 & 16.5 & .59 & 19.8 & .79 & +3.4 \\
 & NAFNet    & 19.8 & .78 & 16.5 & .61 & 19.8 & .78 & +3.3 \\
 & ResUNet   & 19.7 & .79 & 16.5 & .60 & 19.7 & .79 & +3.2 \\
\midrule
\multirow{6}{*}{\rotatebox{90}{\scriptsize OASIS}}
 & Input     & 10.6 & .13 &  4.5 & .01 & -- & -- & +6.1 \\
 & SimpleCNN & 31.9 & .98 & 12.9 & .38 & 31.8 & .98 & +19.0 \\
 & DnCNN     & 32.9 & .99 & 14.0 & .54 & 32.9 & .99 & +18.9 \\
 & UNet      & 33.3 & .99 & 17.6 & .80 & 33.4 & .99 & +15.7 \\
 & NAFNet    & 33.3 & .99 & 17.6 & .79 & 33.3 & .99 & +15.7 \\
 & ResUNet   & 33.4 & .99 & 17.7 & .81 & 33.4 & .99 & +15.7 \\
\midrule
\multirow{6}{*}{\rotatebox{90}{\scriptsize IXI}}
 & Input     &  9.4 & .14 &  5.7 & .02 & -- & -- & +3.7 \\
 & SimpleCNN & 12.9 & .28 &  9.3 & .10 & 13.3 & .35 & +3.6 \\
 & DnCNN     & 15.6 & .45 & 11.0 & .24 & 16.0 & .59 & +4.6 \\
 & UNet      & 18.2 & .70 & 13.3 & .43 & 18.3 & .76 & +4.9 \\
 & NAFNet    & 18.1 & .72 & 13.3 & .45 & 18.2 & .74 & +4.9 \\
 & ResUNet   & 18.2 & .74 & 13.2 & .43 & 18.2 & .76 & +4.9 \\
\midrule
\multirow{6}{*}{\rotatebox{90}{\scriptsize Abdomen}}
 & Input     &  4.6 & .04 &  4.7 & .02 & -- & -- & $-$0.1 \\
 & SimpleCNN & 16.8 & .49 & 12.7 & .16 & 17.2 & .54 & +4.1 \\
 & DnCNN     & 19.4 & .80 & 14.2 & .37 & 20.0 & .81 & +5.3 \\
 & UNet      & 22.4 & .91 & 16.7 & .77 & 22.7 & .91 & +5.6 \\
 & NAFNet    & 21.7 & .85 & 17.0 & .69 & 22.0 & .87 & +4.7 \\
 & ResUNet   & 22.3 & .89 & 16.9 & .75 & 22.6 & .88 & +5.5 \\
\bottomrule
\end{tabular}
\end{table}

The answer (Table~\ref{tab:main}) is not the expected one. After
reconstruction the SQUID leads on every dataset and every model, but
the \emph{margin} swings sharply with the anatomy: the same UNet that
narrows the gap to +3.4\,dB on BrainWeb blows it open to +15.7\,dB on
OASIS. The raw input gap does not predict this. SQUID and
coil start within 3.7--6.1\,dB on the brain datasets and roughly even
on abdomen, yet ML can either compress that lead or amplify it
several-fold. SSIM tells the same story, and the effect is a property of
model capacity rather than of any one network: NAFNet stays within
0.2\,dB of UNet and ResUNet on the brain datasets. Two-channel fusion
pays off only where the sensors are genuinely complementary (a modest
gain on BrainWeb and abdomen) and is redundant on OASIS, where the
SQUID input is already near-perfect (SSIM 0.99).

\begin{figure}[t]
  \centering
  \includegraphics[width=0.88\columnwidth]{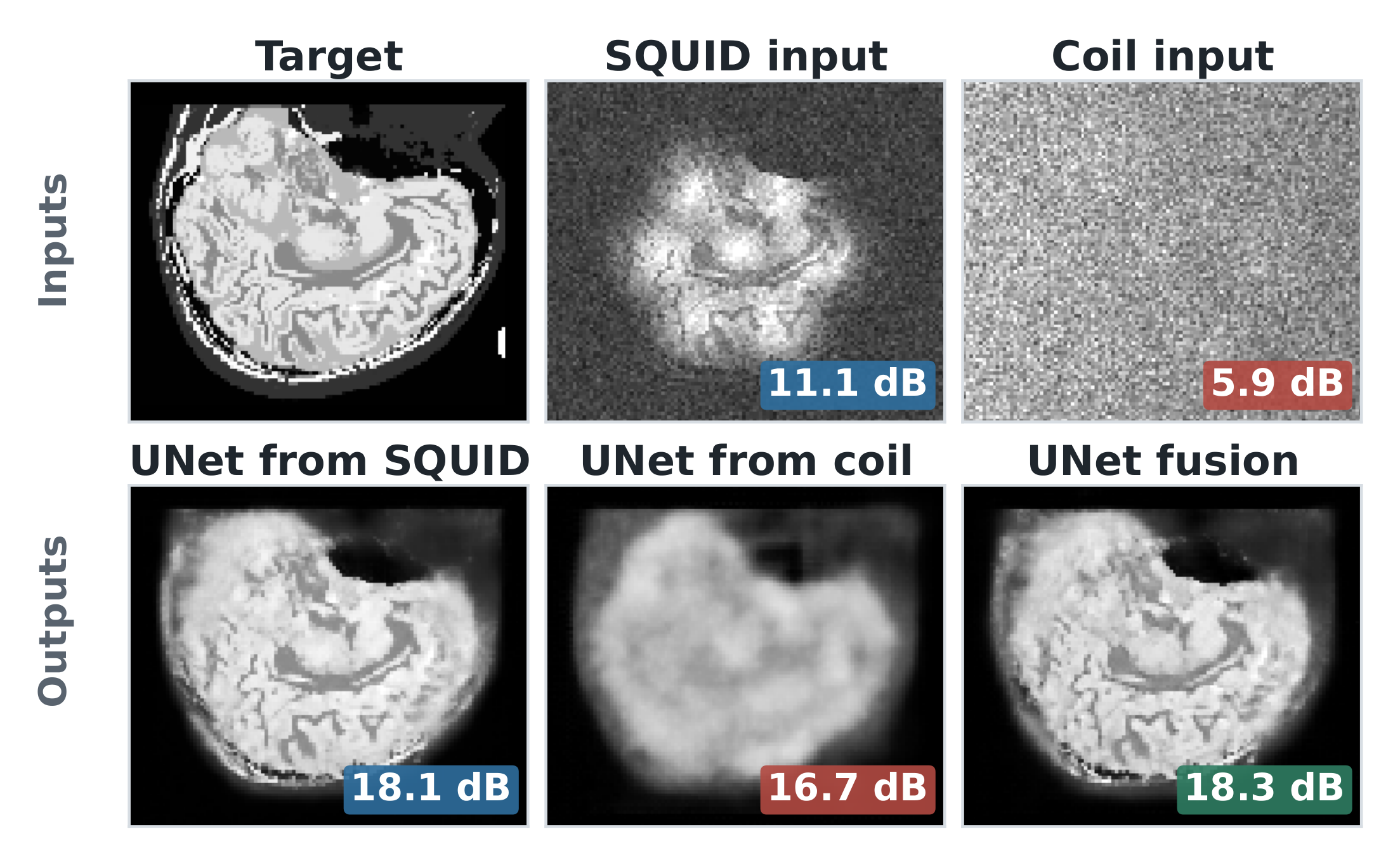}
  \caption{Reconstruction on a single representative BrainWeb test slice (per-slice PSNR shown; test-set means are in Table~\ref{tab:main}). Top row: ideal target and noisy inputs from SQUID and coil. Bottom row: UNet outputs from each sensor configuration.}
  \label{fig:recon}
\end{figure}

Fig.~\ref{fig:recon} shows representative BrainWeb reconstructions.
The SQUID input preserves gross brain anatomy while the coil input is
dominated by noise. After UNet reconstruction, the SQUID-derived result retains finer detail.

\begin{figure}[t]
  \centering
  \includegraphics[width=0.88\columnwidth]{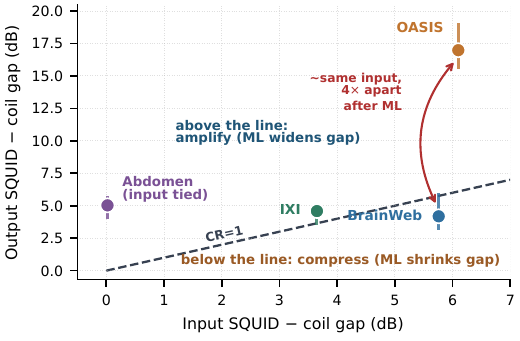}
  \caption{Input vs.\ output SQUID-coil PSNR gap for four datasets (marker: 5-model mean, whisker: min--max across the five networks). The dashed CR\,$=$\,1 line is where ML leaves the gap unchanged; above it the gap is amplified, below compressed.}
  \label{fig:study1}
\end{figure}

To make this precise we define the compression ratio (CR) as the
output gap divided by the input gap: CR\,$=$\,1 means ML preserves the
physical difference, CR\,$<$\,1 means it compensates for the weaker
sensor, and CR\,$>$\,1 means it magnifies the stronger sensor's lead.
The ratio is meaningful only when the sensors differ at the input;
where they start essentially tied (abdomen), we report the absolute dB
change instead.
Fig.~\ref{fig:study1} plots the output gap against the input gap for
each dataset, with the five models shown as a min--max whisker.

The CR makes the dataset dependence concrete, spanning from
compression on BrainWeb (UNet 0.59$\times$, SimpleCNN 1.02$\times$)
through near-preservation on IXI (0.99--1.35$\times$) to strong
amplification on OASIS (UNet 2.57$\times$, SimpleCNN 3.11$\times$). On
abdomen the input gap is near zero, yet ML still creates a +4 to
+6\,dB SQUID advantage.

These behaviors reflect how training-set size and anatomical regularity combine. When data is abundant and regular, the network recovers fine detail from the cleaner SQUID input (OASIS), so the gap widens; when data is limited (BrainWeb), neither sensor yields fine detail and the gap compresses. The abdomen case is the extreme one: the SQUID's distinct spatial sensitivity profile produces an advantage even though the two sensors start essentially tied.

\smallskip\noindent
\textit{Design Rule~1:
The interaction between ML reconstruction and sensor quality is
dataset-dependent. On data-rich problems ML can amplify the SQUID
advantage; on data-limited problems it can compress the gap. System
designers must evaluate sensor choices through end-to-end simulation
rather than relying on SNR alone.}

\subsection{Field-Strength Sweep and Sensor Crossover}
\label{ssec:study2}

ULF spans two orders of magnitude in $B_0$, and the coil's
disadvantage shrinks as its EMF grows with frequency. At what field
does the room-temperature coil become good enough to retire the
cryogenics, and is that threshold a property of the sensors or of the
reconstruction model? We sweep $B_0$ from 30 to 10{,}000\,$\mu$T at
NEX\,=\,50 for three models (SimpleCNN, UNet, ResUNet).

\begin{figure}[t]
  \centering
  \includegraphics[width=0.88\columnwidth]{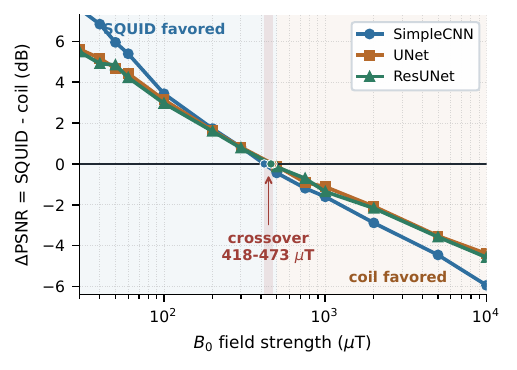}
  \caption{PSNR gap (SQUID minus coil) vs.\ $B_0$ for three models at NEX\,=\,50. Positive values indicate SQUID is the superior sensor; negative values favor the coil. All three models cross zero near 450\,$\mu$T; the shaded band (418--473\,$\mu$T) spans the three models' individual zero-crossings, confirming that the crossover is driven by sensor physics rather than model choice.}
  \label{fig:study2}
\end{figure}

The threshold is real, and it belongs to the sensors, not the network
(Fig.~\ref{fig:study2}). The SQUID-minus-coil gap starts positive and
falls monotonically as the coil EMF grows with $\omega_0 = \gamma B_0$,
crossing zero near 450\,$\mu$T; all three models cross within a narrow
418--473\,$\mu$T band, so the crossover is fixed by Faraday gain, not by
ML capacity. Its location is robust to the number of averages: the
NEX\,=\,50, 70, and 100 sweeps all cross within the same 418--473\,$\mu$T band, which
reinforces that it is set by sensor physics, not acquisition. Above it the
coil, which needs no cryogenics and costs an order of magnitude less,
is the sensor of choice. We run this sweep on BrainWeb, but unlike
Rule~1's magnitude, which is data-bound, the crossover is set by the
sensor transfer functions (Eqs.~\ref{eq:squid}--\ref{eq:coil}), which
carry no anatomical assumptions, so we expect the field threshold to
transfer across anatomies.

\begin{figure}[t]
  \centering
  \includegraphics[width=0.9\columnwidth]{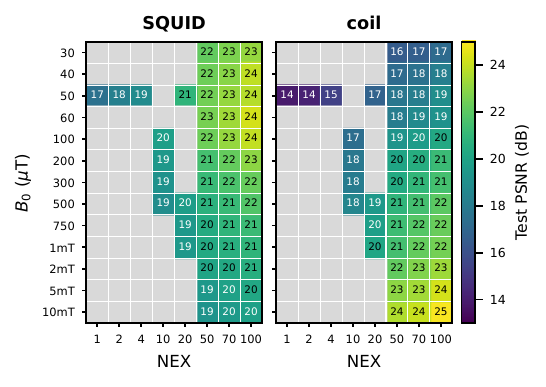}
  \caption{Design-space coverage: test PSNR (UNet) over the swept
    $(B_0, \text{NEX})$ grid for the SQUID (left) and coil (right)
    sensors. Brighter is better; gray cells were not trained,
    exposing the sweep's goal-directed, non-exhaustive coverage. SQUID
    leads at low $B_0$ and the coil at high $B_0$; the crossover of
    Fig.~\ref{fig:study2} is where they switch.}
  \label{fig:landscape}
\end{figure}

Fig.~\ref{fig:landscape} places this crossover in the larger picture:
it shows the reconstruction quality the sweep achieved across the full
$(B_0, \text{NEX})$ grid for each sensor. SQUID dominates the
low-field corner and the coil the high-field corner, and the crossover
is simply where the two panels exchange the lead; the blank
cells expose the sweep's sparse, goal-directed coverage.

\smallskip\noindent
\textit{Design Rule~2:
For the modeled sensor geometries and noise, SQUID is the superior
sensor below roughly $450\,\mu$T; above this field, the room-temperature
induction coil suffices, eliminating the need for cryogenic
infrastructure.}

\subsubsection{NEX--model trade-off}

\begin{figure}[t]
  \centering
  \includegraphics[width=0.9\columnwidth]{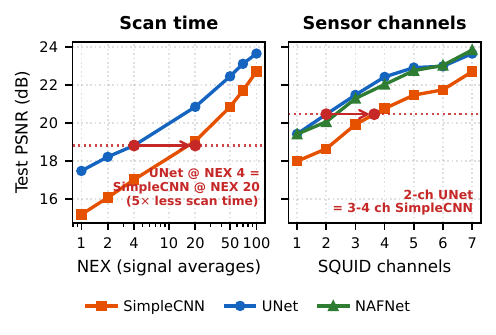}
  \caption{Two faces of hardware--ML substitutability on SQUID
    ($B_0 = 50\,\mu$T). \textbf{Left:} test PSNR vs.\ NEX, where UNet at
    NEX\,=\,4 matches SimpleCNN at NEX\,=\,20, trading model capacity
    for a $5\times$ shorter scan. \textbf{Right:} test PSNR vs.\ SQUID
    channel count, where a 2-channel UNet matches SimpleCNN between its 3- and 4-channel points,
    trading model capacity for sensor channels.}
  \label{fig:tradeoff}
\end{figure}

Acquisition time is the other currency a designer spends on quality:
NEX averaging buys input SNR as $\sqrt{\text{NEX}}$ at a linear cost in
scan time. Can model capacity buy the same quality for less time?
Sweeping NEX from 1 to 100 at $B_0 = 50\,\mu$T, it can: on SQUID, UNet
at NEX\,=\,4 (18.8\,dB) matches SimpleCNN at NEX\,=\,20 (19.0\,dB): a
$5\times$ reduction in scan time bought with a larger network
(Fig.~\ref{fig:tradeoff}, left).

On the coil the substitution is far less efficient: UNet at NEX\,=\,4
reaches 15.3\,dB, while SimpleCNN requires NEX\,=\,50 to reach a
comparable 14.9\,dB, a $12.5\times$ ratio. The gap between SQUID
and coil also widens with NEX: for UNet, from +3.6\,dB at NEX\,=\,1 to
+4.7\,dB at NEX\,=\,100, because additional averaging benefits the
SQUID signal (which contains recoverable anatomical detail) more
than the coil signal (which remains noise-dominated).

These results add a temporal dimension to Rule~2: when scan time is
limited, using a larger model is more effective than increasing NEX,
particularly for SQUID-based systems.

\subsection{Channel Count vs.\ Model Capacity}
\label{ssec:study3}

A SQUID array trades cryogenic channels for image quality; a larger
network trades compute for the same. Are the two resources
interchangeable, and at what exchange rate? We vary the SQUID channel
count from 1 to 7 at $B_0 = 50\,\mu$T and NEX\,=\,100 for three models
(SimpleCNN, UNet, NAFNet).

They are, and the rate is concrete (Fig.~\ref{fig:tradeoff}, right).
Quality rises monotonically with channel count but saturates above
4--5 channels, and model capacity buys the same quality as cryogenic
channels: a 2-channel array with UNet (PSNR 20.5\,dB) matches SimpleCNN
only between its 3- and 4-channel points. A system designer can therefore trade
sensor hardware for ML capacity in either direction, against
constraints such as dewar size, cooling capacity, or available
compute, though the exact exchange rate depends on the model pair and
operating point.

Adding a coil channel provides a fusion gain at every
channel count, confirming that
the coil's complementary $B_1$ spatial profile adds value even
alongside multiple SQUID channels.  The fusion gain is largest at low
channel counts (1.0\,dB at 1 channel for UNet) and decreases as
channels increase (0.4\,dB at 6 channels), because additional SQUID
channels progressively cover 
regions where the coil previously provided unique information.

\smallskip\noindent
\textit{Design Rule~3:
In hardware/software co-design, channel count and model capacity are substitutable: for systems with
strict hardware constraints, a more capable ML model can partially
compensate for fewer SQUID channels. Returns diminish above 4--5 channels.}

\subsection{Discussion}

Rule~1's per-dataset behavior (Section~\ref{ssec:study1}) is itself the key finding: there is no universal rule for how ML interacts with sensor quality. The broader lesson is methodological: once a learned component sits in the signal path, component-level figures of merit such as SNR no longer predict system quality, and end-to-end co-simulation becomes the only reliable predictor. Nothing about that structure (a simulated acquisition front-end and a learned reconstruction back-end joined by a typed IR) is MRI-specific: the receiver base class, HDF5 contract, and idempotent sweep make no MRI assumptions, so any modality with a simulable forward model could reuse the tool chain by supplying a new front-end and dataset generator.

Our conclusions are simulation-based. The physics
is validated for numerical correctness (Table~\ref{tab:validation}) and
calibrated to published measurements~\cite{vesanen2013, zotev2008,
clarke2004}, but hardware imperfections such as patient motion,
environmental magnetic interference, and manufacturing tolerances are
not yet modeled. PSNR and SSIM measure fidelity to the simulated ground
truth rather than task-level quality such as lesion detectability; the
four datasets are brain and abdomen simulants, leaving anatomies such as
cardiac and spine untested; and each configuration is trained with a
single seed. Validating the rules on scanner data (Rules~2 and~3 are
swept on BrainWeb) is the natural next step, though paired ground-truth
ULF-MRI data remains scarce.

\section{Conclusion}
\label{sec:conclusion}

We presented OpenULF, a modular, open-source toolkit for data-driven ultra-low-field MRI with both classical and quantum sensors. Across four datasets and 774 trained models, three explorations yield findings that SNR analysis alone cannot provide: the ML-sensor interaction is dataset-dependent, compressing the SQUID-coil gap on some datasets and amplifying it on others; the SQUID-coil crossover near 450\,$\mu$T is where the room-temperature coil takes over from the cryogenic SQUID; and channel count and model capacity are interchangeable design variables, with diminishing returns above four to five SQUID channels. The OpenULF code is released as open source, with the trained models to follow on publication; because the pipeline is configuration-driven, researchers can substitute their own architectures or datasets to explore questions beyond those reported here. Future work includes validation against real ULF-MRI hardware and support for sensors such as optically pumped magnetometers.

\section*{Acknowledgment}
This work was supported in part by the U.S. National Science Foundation
under Grants 2507948, 2513431, 2440637, and 2612445.

\bibliographystyle{IEEEtran}
\bibliography{refs}

\end{document}